\newcommand{\I}{Isabelle}
\newcommand{\myTitle}{Isabelle Formalisation of Original Representation Theorems}
\newcommand{\restr}[2]{{\left. #1 \right|}_{#2}}

\newcommand{\N}{\mathbb{N}}
\newcommand{\sdiff}{\backslash}
\newcommand{\emp}{\emptyset}

\newcommand{\fpow}[1]{{\overline{2}}^{#1}}

\newcommand{\olap}{\between}

\documentclass
[
]
{llncs}

\usepackage{cancel}
\usepackage{cite}
\usepackage{amsmath,amssymb,amsfonts}
\usepackage{graphicx}
\usepackage[]{hyperref}
\hypersetup{%
pdftitle={\myTitle{}}%
, pdfauthor={Marco B. Caminati}%
, pdfstartview=FitBH
, pdffitwindow=true
, pdfkeywords={models of computation,
semantics,
intelligent mathematics,
AI-aided mathematical discovery,
concurrency,
event structures,
full graphs,
representation theorems, event structures}
, pdflang=en-GB
, pdfpagemode=FullScreen
, pdfstartview= 100 100 10000
, pdfsubject={intelligent mathematics}
, pdfwindowui=true
, colorlinks=true
, linkcolor=blue
, urlcolor=blue
, citecolor=green
, breaklinks = true
, menucolor=brown
}
\usepackage[utf8x]{inputenc}
\usepackage[]{listings}
\usepackage{textcomp}
\usepackage{xcolor}

\title{\myTitle{}%
\author{
\href{http://orcid.org/0000-0002-4529-5442}
{Marco~B.~Caminati}\orcidID{0000-0002-4529-5442}
}
\institute{%
School~of~Computing~and~Communications
\\
Lancaster~University~in~Leipzig
\\
Nikolaistrasse 10
\\
04109 Leipzig
\\
Germany
\\
\email{m.caminati@lancaster.ac.uk}
}
}

\DeclareMathOperator{\dom}{dom}
\DeclareMathOperator{\ran}{ran}
\DeclareMathOperator{\fie}{fie}

\begin{document}
\maketitle{}
\begin{abstract}
In a recent paper, new theorems linking apparently unrelated mathematical objects (event structures from concurrency theory and full graphs arising in computational biology) were discovered by cross-site data mining on huge databases, and building on existing Isabelle-verified event structures enumeration algorithms.
Given the origin and newness of such theorems, their formal verification is particularly desirable.
This paper presents such a verification via Isabelle/HOL definitions and theorems, and exposes the technical challenges found in the process.
The introduced formalisation completes the verification of Isabelle-verified event structure enumeration algorithms into a fully verified framework to link event structures to full graphs. 
\end{abstract}
\section{Introduction}
\label{RefSectIntro}
In~\cite{cicm_de5a57ab1581417ba20bdb26fed8116f}, the first machine-verified contribution to the \emph{Online Encyclopedia of Integer Sequences}  (\href{http://oeis.org/}{\emph{OEIS}})~\cite{sloane2013line} was presented, through an \I{}/HOL-verified algorithm enumerating all labeled \emph{prime event structures} (or just event structures, or even only \emph{ES's}).
In~\cite{mbcMining}, a mining technique over massive sets of documents permitted to unearth unforeseen connections between apparently unrelated mathematical domains. 
One particular connection was, in the same paper, explored, linking event structures (via the algorithm from~\cite{cicm_de5a57ab1581417ba20bdb26fed8116f}) to \emph{full graphs} (\emph{FG}s).
Event structures are originated in the study of concurrent computational systems, while full graphs arise in the field of computational biology~\cite{fulkerson1965incidence}. 
In~\cite{mbcMining}, the deeper motivation of this connection was found as being given rise by a new representation theorem for event structures and a set of derived results, cross-fertilising between the two fields and permitting to obtain new theorems for both the related objects (ES's and FGs).
The two papers \cite{cicm_de5a57ab1581417ba20bdb26fed8116f} and \cite{mbcMining}, therefore, complement each other to provide enumerating algorithms and new connections found using the former. 
However, only the results from~\cite{cicm_de5a57ab1581417ba20bdb26fed8116f} have been mechanically checked.
The present paper completes the work by providing a \I{}/HOL (from now on, just \I{})~\cite{Isabelle-HOL} formalisation of the representation theorem, the theorem connecting ES's and FGs, a number of related \I{}  definitions and tools, and a computable \I{} isomorphism providing the connection between ES's and FGs.

Section~\ref{RefSectStructs} introduces the subjects of the discourse (e.g., event structures and full graphs), Section~\ref{RefSectTheorems} provides the pen-and-paper version of the theorems formalised, 
Section~\ref{RefSectForm} illustrates the main formalised theorems and definitions,
Sections~\ref{RefSectBij} and~\ref{RefSectRepr}, respectively, illustrate the formalisation of the two main theorems,
while 
Section~\ref{RefSectMeta} contains overall consideration about the formalisation process.
Section~\ref{RefSectConclusions} concludes.

\section{Event Structures and Full Graphs}
\label{RefSectStructs}
This section formally introduces the objects of our theorems. 
To make this paper self-contained, it summarises, together with the subsequent one, the main elements of Sections IV and VI of~\cite{mbcMining}.

\subsection{Event Structures}

A prime event structure (or simply event structure, ES) describes a concurrent computation by identifying the computational events that are causally related and those that exclude one another.
According to the following definition, this is achieved via two relations: $\le$ (causality) and $\#$ (conflict).

\begin{definition}
\label{RefDefEs}
An event structure is a pair of relations $ \left( \leq, \# \right)$ where $\leq$ is a partial order, $\#$ is irreflexive and symmetric, 
$ \left( \fie \leq  \right) \supseteq \left(  \fie \#  \right)$ 
is called the set of events, and for any three events $x_0, x_1, y$: $ x_0 \# y \wedge x_0 \leq x_1 \to x_1 \# y$. 
\end{definition}

\begin{figure}[htbp]
\centering{}
\includegraphics[scale=.75]{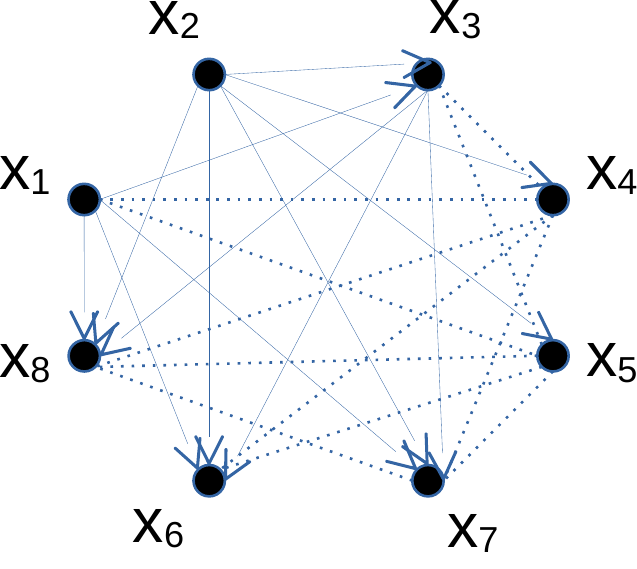}
\caption{An example event structure, with eight events related by causality (denoted by an arrow standing for $\leq$) and conflict (denoted by a dashed line).}
\label{RefFigEs}
\end{figure}

The last condition is referred to as conflict propagation.
In Definition~\ref{RefDefEs}, $\fie$ denotes the field of a relation: that is, the union of its domain ($\dom$) and range ($\ran$).
The usual infix notation for the relations in Definition~\ref{RefDefEs} can become inconvenient, therefore we also introduce an additional notation representing the relations with letters, writing, e.g., $\left( x,y \right) \in D$ instead of $x \leq y$ and $\left( x,y \right) \in U$ in lieu of $x \# y$.
We will typically use the letters $D$ and $U$ as above to suggest the reader what they encode: $D$ stands for ``directed'' and $U$ for ``undirected''.
Indeed, $\leq$, as a partial order, is naturally viewable as a directed graph and $\#$, being symmetric, as an undirected graph. 
See also the comment immediately after Definition~\ref{RefDefFull}.
Since any finite relation is a graph having its vertices (or nodes) coinciding with the field of the relation, and since, for any finite partial order, that graph can be naturally made a directed graph, it is easy to represent any finite ES via diagrams such as the one in Fig.~\ref{RefFigEs}.

\subsection{Full Graphs}
Any family of sets can be used to build a graph where each vertex represents a set of the family, an undirected edge connects overlapping sets, and a directed edge connects a superset to a subset.
Such a construction occurs when studying the following problem: given $n$ subsets of a given set of $m$ elements, is there a way of labeling the elements with natural numbers such that the element occur consecutively (with respect to this labeling) in each subset?
One practical application of this labeling problem arises in bioinformatics, where the elements of subsets represent observed blemishes to parts of a gene, which are supposed to be more likely to affect parts of the gene which are connected: therefore, finding such a labeling can provide essential information about the topology of a gene~\cite{fulkerson1965incidence,benzer1959topology}.
The graphs that can be created in this way are specified by Definition~\ref{RefDefFull}.

\begin{definition}
\label{RefDefFull}
A \emph{full graph} (FG) is a mixed, unweighted, simple%
\footnote{Recall that a graph is \emph{simple} when it has no self loops and no multi-edges; it is \emph{mixed} when it has both directed and undirected edges. See~\cite[Section~1.1]{gross2003handbook}.}
graph over vertices $V$, of directed edges $D$, and undirected edges $T$ such that there is an injective function $f$ on $V$ yielding non-empty sets and with the property
\begin{align}
\forall x, \ y \in V.\ 
\left( 
\left( x, y \right) \in D \leftrightarrow f\ x \supseteq f\ y\  \right) \wedge
\\
\left( 
\left( x, y \right) \in T \leftrightarrow f\ x \text{ and } f\ y \text{ overlap} \right);
\end{align}
here, we say that two sets $A$ and $B$ \emph{overlap} (written $ A \olap B $) when $A \cap B \notin \left\{ A, B, \emp \right\}$.
We call $f$ an \emph{fg-representation} of the full graph $(D,T)$.
Alternatively, we will say that $T$ \emph{makes a full graph of} $D$ (through $f$) when such an fg-representation $f$ exists.
\end{definition}

Having insisted in Definition~\ref{RefDefFull} in encoding $T$ via ordered pairs, even though is an undirected graph, makes that encoding redundant; however, this is convenient because we can then regard $T$ as a (symmetric) relation, as all the other components in the definitions of ES's and FGs, also thanks to the fact that all these components are simple graphs, making the encoding as relations adequate.

\section{Connecting ES's and FGs}
\label{RefSectTheorems}
In~\cite{mbcMining}, a systematic way of looking for matches between entries in the OEIS and free text search results across Google and Google Scholar is introduced, producing thousands of unexplored and potentially interesting matches.
One of them relates the enumerations of ES's (introduced in OEIS by~\cite{cicm_de5a57ab1581417ba20bdb26fed8116f}) and of FGs (in~\cite[Section~4]{cowen1996enumeration}): the number of (labeled) ES's and of FGs over a fixed number of vertices $n$ coincide for small $n$.
In the same paper~\cite{mbcMining}, this connection is explored, motivated and proven by providing a one-to-one map between ES's and FGs, which is an isomorphism once framed as a mapping between representations of ES's and FGs.

\begin{figure}[htbp]
\centering{}
\includegraphics[scale=.75]{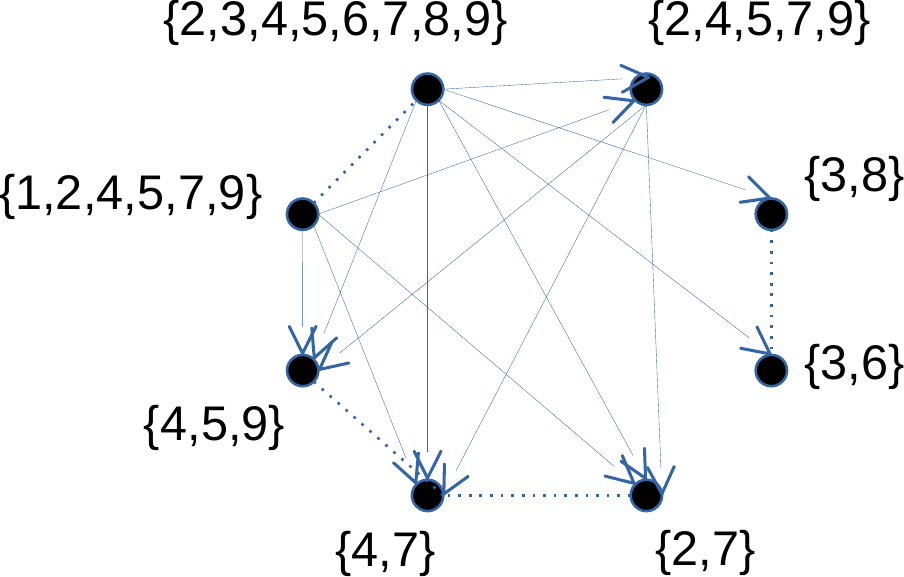}
\caption{The full graph isomorphic to the event structure of Fig.~\ref{RefFigEs}. This is the full graph example originally featured in Section~3 of \cite{fulkerson1965incidence}. Here, the arrows represent $\supseteq$, and the dashed lines the overlapping relation.}
\label{RefFigFg}
\end{figure}

To introduce the ideas in the latter paper, we start by looking at the evaluation, through this isomorphism, of the ES in Fig.~\ref{RefFigEs}, giving the FG in Fig.~\ref{RefFigFg}. 

To make more precise the similarity between the figures, we must understand how they are generated: in Fig.~\ref{RefFigFg}, the edges are determined by looking at operations on sets associated to each node.
In this sense, we have a representation of the FG in terms of set-theoretical notions, indirectly dictating the structure of the FG itself by definition.
In the case of event structures, however, such a representation is absent in the definition, which dictates the property of the structure directly by imposing relationships between $\leq$ and $\#$.
To formally link the connection we are looking at, we must find a representation for the ES as well, 
through a suitable definition of ES-representation and a \emph{representation theorem} establishing that an equivalent definition of ES can be given in terms of such a representation, as done with FGs.
This is an interesting endeavour in general, not limited to the specific task of finding connections between different domains: see~\cite[Section~III]{mbcMining}, which also discusses and details the notion of representation.

The following definition will turn out to yield adequate representations for ES's.

\begin{definition}
\label{RefDefRepr}
Given two binary relations $D$ and $U$, the set-valued function $f$ is a \emph{representation} for $(D,U)$ if
\begin{align}
\label{RefFmReprSub}
\forall x \ y \in \dom f.\ &\left( \left( x, y \right) \in D \leftrightarrow f \left( x \right) \supseteq f \left( y \right) \right) 
\ \wedge \\ 
\forall x \ y \in \dom f.\ 
& \label{RefFmReprDisj}
\left( 
\left( x, y \right) \in U \leftrightarrow f \left( x \right) \cap f \left( y \right) = \emp
\right)
. &
\end{align}
Here, we say that, given $D$ and any $U$ with $\fie U \subseteq \fie D$, any such a $f$ (if it exists) is called \emph{admissible}. 
\end{definition}

And by adequate we mean that the following representation theorem holds.

\begin{theorem}[Representation theorem]
\label{RefLmRepr}
Consider two binary relations $D$ and $U$, with $D$ finite and $\fie U \subseteq \fie D$. Then $\left( D, U \right)$ is an event structure if and only if there is an injective representation $f : \fie D \to \fpow{\N}\sdiff \left\{ \emp \right\}$ for $(D, U)$,
\end{theorem}
where $\fpow{X}$ denotes the finite subsets of $X$.

Our second representation theorem for event structures, Theorem~\ref{RefLmBij}, offers a bijective construction connecting them to full graphs.

\begin{theorem}
\label{RefLmBij}
Consider a finite relation $D$ and a function $F_D$ mapping working as follows on its argument $R$:
\begin{align*}
F_D:=R \mapsto \left( \fie D \times \fie D \right) \sdiff \left( D \cup D^{-1} \right) \sdiff R.
\end{align*}
A bijection between\\
$X:=\left\{ T | T \text{ makes a full graph of } D \right\} $ and\\ 
$Y:=\left\{ U | U \text{ is admissible for } D \right\} $ \\
is given by $\restr {F_D} {X}$.
\end{theorem}

\begin{figure}[h]
\centering{}
\includegraphics[scale=.75]{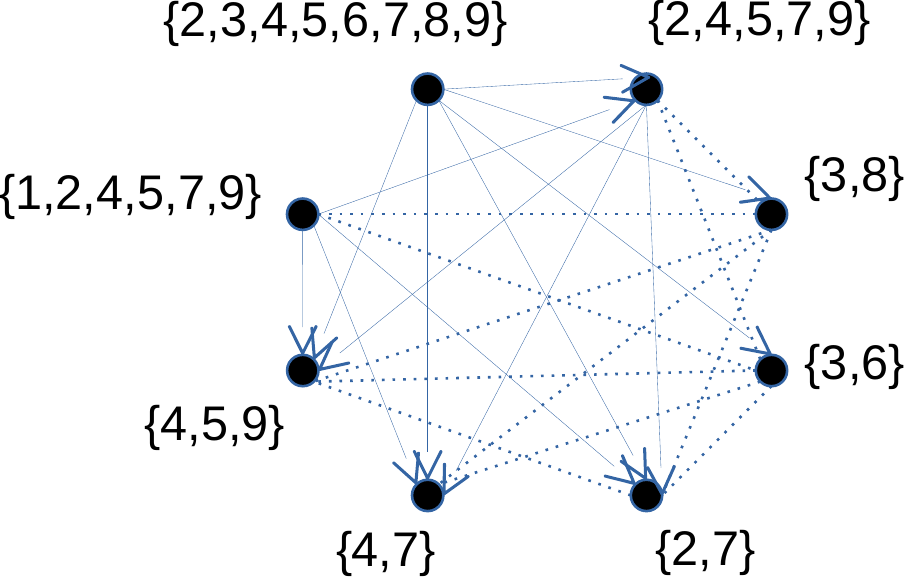}
\caption{A representation for the event structure of Fig.~\ref{RefFigEs}. Now, the arrows represent $\supseteq$ and the dashed lines the disjointness relation. Theorem~\ref{RefLmRepr} states that any set of events is an event structure if and only if such a representation is constructible.}.
\label{RefFigRepr}
\end{figure}

Fig.~\ref{RefFigRepr} attaches a representation (always existing, according to Theorem~\ref{RefLmRepr}) to the ES of Fig.~\ref{RefFigEs}.
Using $F_D$ as in Theorem~\ref{RefFigRepr}, one can now promptly relate that ES to the FG of Fig.~\ref{RefFigFg}.

\section{Formalisation and Verification: Introduction}
\label{RefSectForm}
We start from the top level, that is, the \I{} renditions of the main theorems.

Theorem~\ref{RefLmRepr} is stated as
\begin{lstlisting}[caption=\I{} rendition of Theorem~\ref{RefLmRepr}, label=RefLmReprCode]
theorem representation: assumes "finite D" 
	"Field U ⊆ Field D" shows 
"(isLes D U) = (∃ f. isInjection f & Domain f = Field D & 
	({}::nat set)∉Range f & finite ((Union o Range) f) 
	& isRepresentation f D U)",
\end{lstlisting}

while Theorem~\ref{RefLmBij} reads

\begin{lstlisting}[caption=\I{} rendition of Theorem~\ref{RefLmBij}, label=RefLmBijCode]
theorem bijection: assumes "finite D" 
"F=(λR. ((Field D × Field D) - (D ∪ D^-1)- R))" 
"X={T|T. Field T ⊆ Field D & (∃ f. isInjection f & 
	({}::nat set)∉Range f & Domain f=Field D & 
	isFgRepr f D T)}"
"Y={U|U. Field U ⊆ Field D & (∃ f. isInjection f & 
	({}::nat set)∉Range f & Domain f=Field D & 
	isRepresentation f D U)}" shows 
"F`X=Y & F`Y=X & inj_on F X & inj_on F Y & card X=card Y",
\end{lstlisting}

where \lstinline{inj_on F X} returns true when the total function \lstinline{F} is injective over the set \lstinline{X}, while the notation \lstinline{^-1} denotes the converse of a relation. 

The reader might have noticed a subtle difference between \lstinline{f} occurring in Listing~\ref{RefLmReprCode} and \lstinline{F} occurring in Listing~\ref{RefLmBijCode}: while both are functions, they are implemented very differently within \I{}/HOL.
Indeed, \lstinline{F} is a standard HOL function, a primitive notion in higher order logic~\cite{muller1997treating}; on the other hand, \lstinline{f} is implemented as a set of ordered pairs, in the way standard set theory (e.g., ZF,  Zermelo-Fraenkel set theory~\cite{enderton1977elements}) represents functions.
The verification presented here extensively exploits this duality, choosing one construct or the other depending on the  particular function at hand and on the theorem it appears in.
There are several reasons for this approach: one is that the totality of functions imposed by HOL is sometimes an inconvenience~\cite{muller1997treating} which can be worked around by choosing the second construct; another one is that set theoretical operations on functions, such as union, subtraction, conversion (\lstinline{^-1}) are sometimes useful, and are unavailable with the first construct; as an example of this usefulness, let us take the \lstinline{+<} infix operator, which grows a relation \lstinline{P} with another one \lstinline{Q}, performing overriding if necessary, and is defined as
\begin{lstlisting}
(P - (Domain Q × Range P)) ∪ Q.
\end{lstlisting}

One advantage of this definition is that it works for any pair of relations \lstinline{P} and \lstinline{Q}, and at the same time preserves right-uniqueness if \lstinline{P} and \lstinline{Q} are right-unique (that is, functions).
Additionally, existing facts about the building blocks of \lstinline{+<} (\lstinline{-}, \lstinline{×}, \lstinline{Domain}, \lstinline{Range}, \lstinline{∪}) typically makes proofs about \lstinline{+<} easier, helped by the simplicity of its definition. 
This operator can be conveniently overloaded to the point-wise special case:
\begin{lstlisting}
abbreviation singlepaste where "singlepaste f pair == 
f +< {(fst pair, snd pair)}"
notation singlepaste (infix "+<" 75)
\end{lstlisting}
Note that the type of \lstinline{g} in \lstinline{f+<g} avoids ambiguity for the overloaded \lstinline{+<} operator.

On the other hand, set-theoretical functions are actually relations and, as such, need to be shown to be right unique (by showing they satisfy a dedicated \I{} predicate \lstinline{runiq}) before they can be treated as functions.
Overall, keeping both constructs has the upside of being able to take advantage of the best of both worlds~\cite{caminati2014set2}.

The price to pay for this upside is that we have duplicated versions of most operations on functions, one for each construct. 
For example, if \lstinline{F} is a standard HOL function and \lstinline{f} is a set theoretical function, then the application operation on an argument \lstinline{x} is written \lstinline{F x} for \lstinline{F} and \lstinline{f,,x} for \lstinline{f}; the operation yielding the image of a set \lstinline{X} through the function is \lstinline{F`X} versus \lstinline{f``X}, 
the range operation is \lstinline{range F} versus \lstinline{Range f}; the property of injectivity is \lstinline{inj_on} versus \lstinline{isInjection}, etc. 
Other operations, such as union, intersection, domain, \lstinline{^-1}, restriction (denoted \lstinline{||}), and others, only make sense for set-theoretical functions, although a restriction operating on HOL functions (and denoted \lstinline{|||}) was also introduced. In this case, naturally, the result is a set-theoretical functions, since in HOL all functions are total and cannot therefore be restricted directly~\cite{muller1997treating}.

In practice, the reader needs not to worry about these subtle differences deriving from the duality between HOL functions and set-theoretical functions, which were nevertheless discussed in the 
digression above to prevent confusion.

The first theorem above, in Listing~\ref{RefLmReprCode}, equates the definition of being an event structure (\lstinline{isLes}) to the existence of a representation (whose definition is contained in \lstinline{isRepresentation}), while the second theorem shows that \lstinline{F} (the \I{} rendition of $F_D$ occurring in Theorem~\ref{RefLmBij}) is indeed a bijection between the set \lstinline{Y} of admissible conflicts for \lstinline{D} and the set \lstinline{X} of undirected graphs making \lstinline{D} a full graph.
Since this holds for all finite \lstinline{D}s, we have a verified proof of the mined matches illustrated in Section~\ref{RefSectIntro} and in \cite{mbcMining}.

\lstinline{isLes}, \lstinline{isRepresentation}, \lstinline{isFgRepr} are all straightforward from the pen-and-paper definitions, with the first already used in previous formalisations regarding event structures~\cite{tase_45e8046b98de46aca15b6c4ed5c17a96,cicm_de5a57ab1581417ba20bdb26fed8116f,bowles2019balancing}:

\begin{lstlisting}
definition "isLes causality conflict = 
propagation conflict causality & sym conflict & 
irrefl conflict & trans causality & 
antisym causality & reflex causality",
\end{lstlisting}

\begin{lstlisting}
definition "isRepresentation f D U = ∀x∈Domain f. 
	(∀y∈Domain f. ((((x, y)∈D)=(f,,x ⊇ f,,y)) & 
	(((x,y)∈U) = ((f,,x ∩ f,,y)={}))))"
\end{lstlisting}

\begin{lstlisting}
definition "isFgRepr f D T = ∀x∈Domain f. 
	(∀y∈Domain f. ((((x, y)∈D)=(f,,x ⊇ f,,y)) & 
	(((x,y)∈T) = ((f,,x) overlaps (f,,y)))))",
\end{lstlisting}

with the definition of overlapping also very close to the paper version and taking advantage of the infix notation definition capabilities of \I{}:

\begin{lstlisting}
definition "Overlap X Y = (X ∩ Y ∉ {X, Y, {}})" 
notation "Overlap" ("_ overlaps ")
\end{lstlisting}

Moreover, \lstinline{propagation} is a synonym for the following:

\begin{lstlisting}
definition "isMonotonicOver conflict causality = 
∀ x y. (x,y) ∈ causality ⟶ conflict``{x} ⊆ conflict``{y}",
\end{lstlisting}

while \lstinline{reflex} was introduced as follows:

\begin{lstlisting}
definition "reflex P = refl_on (Field P) P",
\end{lstlisting}
where \lstinline{refl_on A R} returns true when the relation \lstinline{R} is reflexive over a subset \lstinline{A} of its domain and range.



All the other  \I{} objects occurring above are part of \I{}'s standard library.

\section{Formalisation and Verification: Proof Structure for \texttt{bijection}}
\label{RefSectBij}
We start from the second theorem introduced above, which is the simpler of the two, in that it relates full graphs to sets of admissible conflict relations for a given partial order, while the link between ES representations and ES's is provided by \lstinline{representation}.



The idea for the proof is simple: we just note that the definition of fg-representation (Definition~\ref{RefDefFull}) and of event structure representation (Definition~\ref{RefDefRepr}) are very similar, mainly differing by the substitution of the overlapping relation with that of disjointness; therefore, we introduce the following operator to map between them:

\begin{lstlisting}
λR. (unRel' D - R),
\end{lstlisting}
where the helper \lstinline{unRel'} takes the complement of a relation:
\begin{lstlisting}
abbreviation "unRel' D==(Field D × Field D) - (D ∪ D^-1)".
\end{lstlisting}

Now, the idea is to show that we can pass from event structures to full graphs by applying the above operator to the conflict relation.
To show that, it suffices to show that the set of valid undirected edges for a given \lstinline{D} can be obtained from the set of valid conflict relations for \lstinline{D} by applying the operator above: this is exactly the thesis 
\lstinline{F`X=Y & F`Y=X} appearing in the \lstinline{bijection} theorem's thesis.
By bijectivity, it suffices to show the weaker relations \lstinline{F`X ⊆ F`Y} and \lstinline{F`Y  ⊆ F`X}, which is done by \lstinline{l53a} and \lstinline{l53b} below, respectively:

\begin{lstlisting}
lemma l53a: assumes "F=(λR. (unRel' D - R))" shows 
"F`{T|T. Field T ⊆ Field D & (∃ f. isInjection f 
	& ({}::nat set)∉Range f & Domain f=Field D 
	& isFgRepr f D T)} ⊆ 
{U|U. Field U ⊆ Field D & (∃ f. isInjection f & 
	({}::nat set)∉Range f & Domain f=Field D 
	& isRepresentation f D U)}"
\end{lstlisting}

\begin{lstlisting}
lemma l53b: assumes "F=(λR. (unRel' D - R))" shows 
"F`{U|U. Field U ⊆ Field D & (∃ f. isInjection f & 
	({}::nat set)∉Range f & Domain f=Field D & 
	isRepresentation f D U)} ⊆
{T|T. Field T ⊆ Field D & (∃ f. isInjection f & 
	({}::nat set)∉Range f & Domain f=Field D & 
	isFgRepr f D T)}"
\end{lstlisting}

\lstinline{l53a} and \lstinline{l53b} are sufficient to draw the thesis of \lstinline{bijection} thanks to the following general propositions (the latter provided by \I{}'s standard library):

\begin{lstlisting}
proposition l52a: assumes "finite (X ∪ Y)" "inj_on f X" 
"inj_on f Y" "f`X ⊆ Y" "f`Y ⊆ X" shows "f`X=Y & f`Y=X" 
\end{lstlisting}

\begin{lstlisting}
lemma card_image:
  assumes "inj_on f A"
  shows "card (f ` A) = card A"
\end{lstlisting}

Finally, the hypotheses \lstinline{inj_on f X} and \lstinline{inj_on f Y} can be deduced when \lstinline{X} and \lstinline{Y} are, respectively, the sets appearing in \lstinline{l53b} by another general result:

\begin{lstlisting}
proposition l55: "inj_on (λX. Y-X) (Pow Y)"
\end{lstlisting}
(where \lstinline{Pow} takes the power set), which applies when X and Y take the particular values above thanks to

\begin{lstlisting}
lemma l54bb: assumes "isFgRepr f D T" "Domain f = Field D" 
"Field T ⊆ Field D" shows "T ⊆ (Field D × Field D)-(D∪D^-1)"
\end{lstlisting}

and

\begin{lstlisting}
lemma l54aa: assumes "isRepresentation f D U" 
"({}::nat set)∉Range f" "runiq f" "Domain f = Field D" 
"Field U ⊆ Field D" shows "U ⊆ (Field D × Field D)-(D∪D^-1)",
\end{lstlisting}

where the \lstinline{runiq} predicate was introduced in the discussion after Listing~\ref{RefLmBijCode}.


\section{Formalisation and Verification: Proof Structure for \texttt{representation}}
\label{RefSectRepr}
The proof is in the two directions; that is, having a representation implies being an event structure (theorem \lstinline{main1}) and being an event structure implies having a representation (theorem \lstinline{main2}):

\begin{lstlisting}
theorem main1: assumes "runiq f" 
"Field D ∪ Field U ⊆ Domain f" 
"isRepresentation' f D U" 
	shows 
"isPreorder D & isMonotonicOver U D & sym U & 
(luniq f ⟶ antisym D) & ({}∉(Range f) ⟶ irrefl U)"
\end{lstlisting}

\begin{lstlisting}
theorem main2: assumes "finite D" "isLes D U" obtains 
f::"('a × nat set)set" where "Domain f=Field D & 
	isInjection f & {}∉Range f & 
	finite ((Union o Range) f) & isRepresentation f D U"
\end{lstlisting}

\subsection{Proof of \texttt{main2}}
\label{RefSectionMain2}
The proof for \lstinline{main2} is arguably among the most complex in the project, since it needs to provide a representation for any given ES.
It is done by induction on the cardinality of \lstinline{D}, starting with the base case which can be proved by Sledgehammer~\cite{blanchette2013extending}:

\begin{lstlisting}
proposition ll50a: assumes "f={}" "D={}" shows 
"isRepresentation f D U & Domain f=Field D & 
isInjection f & runiq f & {}∉Range f"
\end{lstlisting}

The induction step now requires to somehow 
pass from a representation \lstinline{f} of a \lstinline{D'} smaller than a given \lstinline{D} to a representation for \lstinline{D} itself. 
This requires to determine two things:
\begin{enumerate}
\item
\label{RefEnumSmaller}
in which sense \lstinline{D'} is smaller than \lstinline{D};
\item
\label{RefEnumNewrepr}
how to construct the new representation from \lstinline{f}.
\end{enumerate}
For~\eqref{RefEnumSmaller}, we set \lstinline{D'} and \lstinline{D} to differ by exactly one \emph{terminal} event: that is, \lstinline{D'} is obtained from \lstinline{D} by removing one event \lstinline{s} with no children in \lstinline{D}.

For~\eqref{RefEnumNewrepr}, we obtain the new representation for \lstinline{D} by just growing \lstinline{f} with one new set \lstinline{RA} representing \lstinline{s}; this growth is done by the \lstinline{+<} operator seen in Section~\ref{RefSectForm}. 
Note that this growth does not affect the values \lstinline{f} has on the old events.
Theorem \lstinline{extension2} below does exactly that, showing that the function resulting from the \lstinline{+<} operation is still a representation for \lstinline{D}.
However, for this thesis to hold, there are three fundamental requirements on \lstinline{RA}, the set representing the new event \lstinline{s}; these requirements must hold for any existing event \lstinline{x}, and appear in the hypotheses of \lstinline{extension2} labeled as \lstinline{hypOverlap}, \lstinline{hypCausality} and \lstinline{hypConflict}. 
The remaining hypotheses are merely technical, expressing obvious requirements such as \lstinline{f} needing to be a function, \lstinline{s} having no children, \lstinline{s} being fresh, etc.

\begin{lstlisting}
theorem extension2: assumes "runiq f" "(s,s)∈D" 
"D``{s}⊆{s}" "s∉Domain f" assumes 
hypOverlap: "∀x∈Domain f. ¬(f,,,x ⊆ RA)" assumes
hypCausality: "∀x∈Domain f. RA ⊆ f,,,x = (x∈D^-1``{s}-{s})" 
assumes 
hypConflict: "∀x∈Domain f. ((f,,,x)∩RA={})=(x∈U^-1``{s})" 
"∀x∈Domain f. ((x,s)∈U) = ((s,x)∈U)" 
"isRepresentation f (D---s s) (U---s s)"
"F=f+<(s,RA)" "RA≠{}" "(s,s)∉U" 
	shows 
"isRepresentation' F D U"
\end{lstlisting}

\lstinline{extension2} presents a couple of new constructs: first, the operator \lstinline{---} allows to remove a pair from a relation, so that, in this case, \lstinline{D} and \lstinline{U} are extensions of \lstinline{D---s s} and \lstinline{U---s s}.
Secondly, the operator \lstinline{,,,} is very similar to \lstinline{,,} seen in Section~\ref{RefSectForm}, but with a slightly more general definition which is technically more convenient in some cases.
Let us start with the definition of \lstinline{---}:

\begin{lstlisting}
definition "bouthside P X Y = 
	P - ((X×Range P) ∪ ((Domain P)×Y))"
notation "bouthside" ("_\\")
definition "singlebouthside P x y = bouthside P {x} {y}" 
notation "singlebouthside" ("_---")
\end{lstlisting}

This definition uses a special case of \lstinline{\\}, which merely removes portions of domain and range from any relation using elementary set-theoretical operations.


\lstinline{extension2} is what we need to obtain our representation theorem. 
However, as we mentioned above, it dictates three conditions on \lstinline{RA} (\lstinline{hypOverlap}, \lstinline{hypCausality} and \lstinline{hypConflict}) for its validity.
We therefore need to build a set \lstinline{RA} satisfying them.
The following result, one of the most technical, builds a suitable \lstinline{RA}, by 
transforming the representation \lstinline|f| occurring in \lstinline|extension2| into an intermediate representation \lstinline|g| before inducting.

\begin{lstlisting}
lemma l46: assumes "isRepresentation f (D---s s) (U---s s)" 
"runiq f & D``{s}={s} & sym U & 
(let dm=Domain in let R=Range in {}∉R f & 
finite ((Union o R) f) & (Domain D)-{s} ⊆ dm f &
	(let d=D---s s in dm f ⊆ Range d & trans d))"
"let d=D---s s in let sparents=d^-1``(D^-1``{s}) in 
let sconfl=U^-1``{s} in 
let sconcurs=Range d-(sparents ∪ sconfl) in 
  finite sconcurs & sconcurs⊆fixPts D & 
  sparents=D^-1``{s}-{s} & 
  irrefl (U||(sconcurs ∪ Domain f)) & 
  sconfl ∩ D^-1``{s}={} & 
  d``sconfl⊆sconfl & 
  isMonotonicOver U (D|^(D^-1``{s} ∪ (Range d - sconfl)))"
shows 
  "∃ l. let N=Max ((Union o Range) f)+1+size l in 
  let d=Domain in let R=Range in 
  let RA=(Union o set)((map (Union o R) l)@[{N}]) in 
  let g=foldl pointUnion f (l@[(D^-1``{s}-{s})×{{N}}]) in 
  let h=g+<(s,RA) in d g=d f & d h=d f∪{s} & {}∉R g & 
  {}∉R h & isRepresentation g (D---s s) (U---s s) & 
  isRepresentation h D U & runiq g & runiq h & 
  (Union o R) h ⊆ {0..<1+N} & 
  (luniq f ⟶ (isInjection g & isInjection h))"
\end{lstlisting}

Although harder to read than \lstinline{extension2}, \lstinline{l46} has the advantage of having moved all the requirements on \lstinline{RA} back to the given event structure $\left( D, U \right)$.
This comes at the price of passing through \lstinline|g|, which is obtained from \lstinline|f| by repeatedly applying the following operator \lstinline|pointUnion| to the given \lstinline|f| over a suitable list of sets, through the standard functor \lstinline|foldl|:

\begin{lstlisting}
definition "pointUnion ff A =
	ff +< ((λx. ff,,,x ∪ A,,,x)|||(Domain A))".
\end{lstlisting}

Recall that \lstinline!|||! is the restriction operator, see Section~\ref{RefSectForm}.

\subsection{Proof of \texttt{main1}}
The proof of theorem \lstinline{main1} is less technical, and is nicely broken into sublemmas each providing a part of the thesis.
The following lemma takes care of the transitivity:
\begin{lstlisting}
lemma l49a: assumes "runiq f" "Field D ⊆ Domain f" 
"∀x0∈Domain f. (∀x1∈Domain f. (((x0, x1)∈D)=(f,,x0⊇f,,x1)))" 
shows "Field D ⊆ fixPts D & trans D",
\end{lstlisting}
(where \lstinline|definition "fixPts P=Domain(Id∩P)"|), 
while this other proposition takes care of conflict propagation:
 
\begin{lstlisting}
proposition l49bb: assumes "Field D ∪ Range U ⊆ Domain f" 
"∀x∈Domain f. (∀y∈Domain f. ((((x, y)∈D)⟶(f,,x ⊇ f,,y)) & 
	(((x,y)∈U) = ((f,,x ∩ f,,y)={}))))"
shows "isMonotonicOver U D"
\end{lstlisting}

The reflexivity is then granted by combining \lstinline|l49a| with this simple but useful fact:

\begin{lstlisting}
proposition l45e: "(∀x∈Field P. (x,x)∈P)=reflex P".
\end{lstlisting}

When writing the formalisation, a guiding principle was to always try to derive particular results from weaker results (whether the latter already exist in some library or not) applicable to more general objects, which can be strengthened to be applied to more particular objects needed in the specific formalisation one is carrying out.
This resulted in over $300$ lemmas, propositions and theorems, and around $50$ new objects defined.

\I{} was also used to work out minimal requirements for particular results. 
For example, in \lstinline|main1|, no finiteness is required over \lstinline|D|, and the particular irreflexivity property is explicitly bound to the additional requirement of \lstinline|f| not yielding the empty set as a representation.
Similarly, in theorem \lstinline|main1| the antisymmetry property of event structures is linked to the representation being an injection.
These details add proof-theoretical information to any development, and are usually hard to keep track of manually with a pen-and-paper proof.

\section{The Formalisation Process}
\label{RefSectMeta}
The code is available 
at%
\footnote{The link requires a reasonably recent browser.}
\url{https://gitlab.com/users/mbc8/contributed}.
The formalisation of the mathematical objects and results introduced above is roughly 2.7kSLOC and 151Kb (36Kb gzipped) of \I{} code; a bit more due to spawned additions to the theories created for event structures for previous papers such as \cite{tase_45e8046b98de46aca15b6c4ed5c17a96,cicm_de5a57ab1581417ba20bdb26fed8116f,bowles2019balancing}.
To quantitatively assess the formalisation, the length of the mathematical parts appearing in~\cite{mbcMining} was computed by converting the relevant pdf to text, obtaining 21502 bytes (8229 gzipped) as a result.
This gives an apparent de Bruijn factor of 7, and an intrinsic one of 4.3. 
There are about 4 pages of mathematical content in~\cite{mbcMining}, whereas the time spent to formalise it has been estimated in around two weeks of work, giving a formalisation cost of 0.5 weeks per page.
All these numerical parameters are approximate, but help giving an idea of the process itself~\cite{asperti2010some,naumowicz2006example}.
It should also be noted that, although \I{}/HOL implementations of graph theory abound (\cite{noschinski2012proof,koutsoukou2023formalisation,noschinski2015graph,nordhoff2012dijkstra}), the present formalisation used none of them, for two reasons: first, although the theorems relate event structures and full graphs, they don't really need much graph theory. Not even basic notions as walks, paths, etc. are even mentioned.
Secondly, our formalisation deals with mixed graphs (i.e., having both directed and undirected edges), thus restricting the available libraries.
The theorem \lstinline{representation} uses 141 facts (including lemmas, propositions, theorems and definitions) included in the file \lstinline{fullGraph.thy}.
The proofs can be divided into automatically generated ones and one with an explicit Isar proof (starting with the \lstinline{proof} keyword). 
A minority of those explicit proofs were generated by Sledgehammer's \lstinline{isar_proof} feature, but most of them were manually written.
In general, the preference is to have small general facts with simple, usually automatic proofs, which are then put together for the more complex, manual proofs.
This yields a proliferation of lemmas which are hopefully reusable.
This approach goes hand-in-hand with the one providing definitions built in blocks on top of more general definitions.
For example, \lstinline{pointUnion} is defined in terms of \lstinline{|||} and \lstinline{+<}, which are in turn defined in terms of elementary set theoretical operations (cartesian product, union, intersection, set difference, etc.).
One of the longest proof is that of \lstinline{l46} (see Section~\ref{RefSectionMain2}), which is 139 lines and about 10Kb.
About 10 results have proof longer than 20 lines, usually substantially longer, and a number of them has to do with the problem of suitably constructing \lstinline{RA} using a reiterated (via \lstinline{foldl}) \lstinline{pointUnion} operation (see Section~\ref{RefSectionMain2}).
Most proofs are non-constructive; for example, they do not provide an algorithm to build representations. 
However, the operator \lstinline{F} appearing in theorem \lstinline{bijection} and allowing to pass from representations to fg-representations and vice-versa is computable.

\section{Conclusions}
\label{RefSectConclusions}
This paper has presented a rare instance of original theorems having been formalised natively: they were born formalised.
More than that, they were discovered thanks to existing formalisations.
Such theorems provide new representations for event structures and unexpectedly link the latter to the unrelated field of computational biology through the notion of full graphs.
This permits to apply results from one domain to another to immediately obtain new theorems (some such  examples are reported in~\cite{mbcMining})
Therefore, an obvious idea for future work is to formally verify these new theorems, which would imply a formalisation for the domain which is currently not formalised: that of full graphs.
Indeed, while event structures have now a reasonable amount of results formalised, no formalisation exists for more advanced results applicable to full graphs, for example those in \cite{kleitman1995asymptotic,cowen1996enumeration}.

Looking at automated theorem proving, the origin of the presented results (obtained via data mining, as explained in~\cite{mbcMining}), can provide avenues to both develop new techniques and test existing ones: thousands of potentially interesting matches similar to the one giving rise to the results presented here were found.

Another future work direction will seek the generalisation of the original theorems presented here: one natural idea is the extension of Theorem~\ref{RefLmRepr} to infinite event structures, which is comparable to how Priestley's representation theorem generalises (in a by no means trivial manner!) Birkhoff's~\cite[Theorem~11.23]{davey2002introduction}.

\bibliographystyle{splncs04}
\bibliography{mbc}
\end{document}